\documentclass[twocolumn]{article}
\usepackage{arxiv}
\usepackage{amsmath,amssymb,amsfonts}
\usepackage[linesnumbered,ruled]{algorithm2e}
\usepackage{graphicx}
\usepackage{comment}
\usepackage{bbm}
\usepackage{booktabs}
\usepackage[textwidth=2cm]{todonotes}
\usepackage{tikz}
\usepackage{abstract}
\usepackage{xcolor} 
\definecolor{dodgerblue}{RGB}{30,144,255}

\newcommand{\vect}[1]{\boldsymbol{\mathbf{#1}}}

\newcommand*{\qrx}{\dot{Q}_\mathrm{RX}}

\newcommand*{\rhoext}{\rho_\mathrm{ext}}
\newcommand*{\tco}{T_\mathrm{c,out}}
\newcommand*{\tci}{T_\mathrm{c,in}}

\newcommand*{\tso}{T_\mathrm{s,out}}

\newcommand*{\tsi}{T_\mathrm{s,in}}

\newcommand*{\dpp}{\Delta P_\mathrm{p}}
\newcommand*{\dps}{\Delta P_\mathrm{s}}

\newcommand{\rv}[1]{#1}

\begin{document}

\title{A Safe Reinforcement Learning Algorithm for Supervisory Control of Power Plants}
\author{Yixuan Sun\\
Mathematics and Computer Science Division\\
Argonne National Laboratory\\
\And
Sami Khairy\thanks{Sami conducted a majority of the work while he was affiliated with the Mathematics and Computer Science Division at Argonne National Laboratory.}\\
Microsoft\\
\And
Richard B. Vilim\\
Nuclear Science and Engineering Division\\
Argonne National Laboratory
\And
Rui Hu\\
Nuclear Science and Engineering Division\\
Argonne National Laboratory
\And
Akshay J. Dave\thanks{Corresponding Author, ajd@anl.gov}\\
Nuclear Science and Engineering Division\\
Argonne National Laboratory}
\date{}

\twocolumn[
\begin{@twocolumnfalse}
\maketitle
    
\begin{abstract}
\normalsize
\vspace{12pt}
Traditional control theory-based methods require tailored engineering for each system and constant fine-tuning. \rv{In power plant control, one often needs to obtain a precise representation of the system dynamics and carefully design the control scheme accordingly.} Model-free Reinforcement learning (RL) has emerged as a promising solution for control tasks due to its ability to learn from trial-and-error interactions with the environment. \rv{It eliminates the need for explicitly modeling the environment's dynamics, which is potentially inaccurate.}
\rv{However, the direct imposition of state constraints in power plant control raises challenges for standard RL methods.} 
To address this, we propose a chance-constrained RL algorithm based on Proximal Policy Optimization for supervisory control. 
Our method employs Lagrangian relaxation to convert the constrained optimization problem into an unconstrained objective, where trainable Lagrange multipliers enforce the state constraints. 
Our approach achieves the smallest distance of violation and violation rate in a load-follow maneuver for an advanced Nuclear Power Plant design.

\vspace{12pt}
\end{abstract}

\end{@twocolumnfalse}
]
\saythanks

\section{Introduction}
\label{sec:introduction}
Traditionally, control systems have relied on classical control theory and feedback control loops, which can be complex and require significant manual engineering~\cite{KARANAYIL20181245}. 
In particular, power plant control requires accurate system dynamics models to ensure safety, efficiency, and reliability. 
The construction of these models requires considerable effort.
Recently, there has been a growing interest in using machine learning and artificial intelligence techniques for power plant control, including RL~\cite{park2022control, sun2021machine}. 
RL methods can potentially reduce the need for manual engineering and improve the performance, safety, and efficiency of power plant operations by allowing control systems to learn from data and adapt to changing operating conditions more efficiently~\cite{fu2018risk, qian2022development, lin2020deep}. 
These previous studies focused on training an agent to perform tasks, which did not consider the operational constraints of states.

Meeting constraints allows the unit to operate optimally by reducing excessive wear \& tear, thus, making the unit more cost-efficient. 
To learn an optimal control policy that incorporates state-level constraints, in this work, we formulate the ``Safe RL'' \cite{brunkeSafeLearningRobotics2021} problem as a chance-constrained optimization problem and use Lagrangian relaxation to convert the problem to an unconstrained setting.
Moreover, we adopt the state-of-the-art policy gradient method to train an intelligent agent to control an advanced Nuclear Power Plant (NPP) during routine operational transients.

The contributions of this work are four-fold: 
\begin{enumerate}
	\item We create a physics-based learning environment for training an intelligent agent using reinforcement learning to control a nuclear power plant. The learning environment is a reduced-order SINDYc~\cite{brunton2016discovering} model, significantly reducing the computational time to obtain environment feedback.
	\item State-level constraints are included in learning the optimal policy, where the problem is converted to an unconstrained setting with Lagrangian relaxation, and the corresponding Lagrange multipliers are trainable parameters.
	\item We propose a new learning scheme where we update the policy and value networks on a faster time scale and learn the Lagrange multipliers on a slower time scale. The faster time scale allows agents to sufficiently learn a policy to follow the load demand under a fixed level of constraint awareness before a necessary increase in the constraint penalty.
	\item By learning the Lagrangians, the proposed model shows the best performance in controlling an NPP during demanding power transients (up to 50 \% reduction in total power for a load-follow).
\end{enumerate}

\section{Background and Related Work}

\subsection{Reinforcement Learning}
RL is a subfield of machine learning that has gained significant attention in recent years due to its potential to enable intelligent agents to learn and adapt to new environments and tasks. At its core, reinforcement learning involves an agent interacting with an environment. The agent takes actions based on its current state and receives feedback through rewards or penalties. Optimizing its actions to maximize its cumulative reward over time allows the agent to make better decisions and achieve its objectives more effectively~\cite{sutton2018reinforcement}. RL has been the alternative to classic control methods due to its flexibility and no need for much manual engineering and has been successfully applied to various domains, including robotics~\cite{kober2013reinforcement, Ibarz_2021}, game play~\cite{mnih2013playing, silver2017mastering}, and natural language processing~\cite{uc2023survey}, and has shown promise in addressing complex real-world problems. 

The RL problem can be mathematically formulated as a Markov Decision Process~(MDP). An MDP consists of a state space, $\mathcal{S}$, action space, $\mathcal{A}$,  rewards, $\mathcal{R}$, process dynamics determined by the state-transition probability, $\mathcal{P}$, and a discount factor, $\gamma$.  
The objective is to maximize an agent's expected cumulative reward~(return) while navigating through an MDP. 
Under a policy, $\pi$, the state value function describes such expected return, $v_{\pi}(s) = \mathbb{E}[\sum_{k=0}^\infty \gamma^{k} R_{t+k+1} | S_t = s]$. The Bellman equation provides a recursive decomposition of the problem.

\begin{equation}\label{eq:bellman}
	v(s_t) =\sum_{a\in \mathcal{A}} \pi(a|s)\sum_{s' \in \mathcal{S},r \in \mathcal{R}}p(s',r|s,a)[ r_t + v(s')],
\end{equation}
where $s$ is the current state and $s'$ is the next state. We aim to find a policy function, $\pi^*$, that maximizes the value function such that

\begin{equation}
	v^*(s) = \max_{\pi} v_{\pi}(s) = v_{\pi^*}(s).
\end{equation}

In this work, we adopt policy gradient methods. Unlike methods that deduce a policy from state-action values, policy gradient methods aim to directly learn a policy $\pi$ using a parametrized model. This approach is particularly useful for problems with continuous action spaces as it allows the model to provide a probability distribution for actions. Furthermore, policy gradient methods are often more sample-efficient than other methods, making them well-suited for our problem setting.

\subsection{Constrained Deep Reinforcement Learning}
One of the primary motivations for deep RL with restricted safety is the increasing deployment of RL agents in safety-critical applications such as autonomous driving, robotics, and healthcare care~\cite{kiran2021deep, khan2020systematic, yadav2021smart}. These applications require that agents operate safely and avoid actions that could cause harm to humans or the environment. However, traditional RL algorithms are often designed to optimize a single objective, such as maximizing a reward signal, without explicitly considering safety. Safe RL seeks to overcome this limitation by integrating safety constraints into the RL framework and designing algorithms that balance safety and performance objectives. This involves developing methods for modeling and predicting safety hazards, enforcing safety constraints during training and testing, and ensuring that the agents' behavior is interpretable and understandable to humans. Safe RL has the potential to revolutionize many industries by enabling the safe and reliable deployment of autonomous systems.

In the MDP framework, there are two significant categories of Safe RL~\cite{garciaComprehensiveSurveySafe}, modifying the optimality criterion and changing the exploration process based on the guidance of a risk metric. The general idea behind modifying the optimization criterion and the reward function is to consider the risk in the objective function RL models tend to optimize. Additional criteria can include worst-case, risk-sensitive, and constrained criteria~\cite{heger1994consideration, patek2001terminating, tamar2012policy}. On the other hand, with safe exploration, the agent seeks to follow a safety-aware exploration and exploitation scheme by incorporating external knowledge, such as teacher advice~\cite{Korupolu2011BeyondR}, or follow a risk-directed exploration~\cite{gehring2013smart}.

Another way of dealing with safe RL problems is to formulate a constrained Markov Decision Process~(cMDP)~\cite{altman1999constrained}. In a cMDP,  the agent must satisfy constraints while maximizing the cumulative reward. In addition to the elements of an MDP, cMDPs contain a constraint function $\mathcal{C}$, specifying the cost or penalty for transitioning from one state to another by taking an action. The constraints are known a priori, and linear programming and Lagrangian methods are commonly used to solve cMDPs~\cite{brunkeSafeLearningRobotics2021}. In our work, we formulate the constraint function as a chance constraint and adopt Lagrangian relaxation to convert it into an unconstrained learning problem. Section~\ref{sec:method} describes the problem setting and formulation details.

\subsection{Power Plant Control (constrained continuous control)}

Controlling power plants is challenging due to the complexity of coordinating multiple systems, such as boilers, turbines, generators, and auxiliary systems.
Power plants must balance the electricity supply with its \textit{demand} to ensure the electrical grid's stability.
The electricity demand varies depending on weather patterns, the availability of competing plants, and the energy demand.
Nuclear power plants (NPPs) have additional layers of complexity to prevent accidents.
Multiple safety systems address the safety risks that have led to NPPs being the safest baseload energy generators \cite{mccombie2016renewable}.
However, NPPs also have high Operation \& Maintenance (O\&M) costs.
The O\&M costs for existing plants are almost an order of magnitude higher than those of a conventional combined-cycle gas turbine plant \cite{us_eia_2022}.

A proposal to address the high O\&M costs has been to shift the operation of NPPs from a baseload operation paradigm to a ``load-following'' one \cite{jenkins2018benefits}.
In load-following, the operators would adjust the plant's electrical output to meet requested changes in load from the grid.
To achieve this, the total thermal energy generated by the plant must be reduced or increased.
In an NPP, this is primarily achieved by controlling the reactivity of the nuclear core by inserting or removing control blades.
Because NPPs are underactuated, i.e., the number of actuators is lower than the number of states, some states vary between high and low power levels.
There are no additional actuators available to constrain the deviation of such states.
Constraining the deviation of such states is essential to optimize the plant's performance.
For example, if an excessive deviation in temperature occurs during routine load-follows, the cyclical thermal stresses can fail components \cite{viswanathan2000failure}.
Therefore, an additional control algorithm is needed to constrain the changes in power levels to ensure that the plant operates optimally.
This work proposes to develop and demonstrate an RL algorithm to address this task.

\section{Supervisory NPP Control}

This section describes a physics-based environment designed for training RL agents that provide supervisory control of an advanced NPP.

\subsection{RL Environment Design}

\begin{figure}[!ht]
\centering
   \includegraphics[width=\columnwidth]{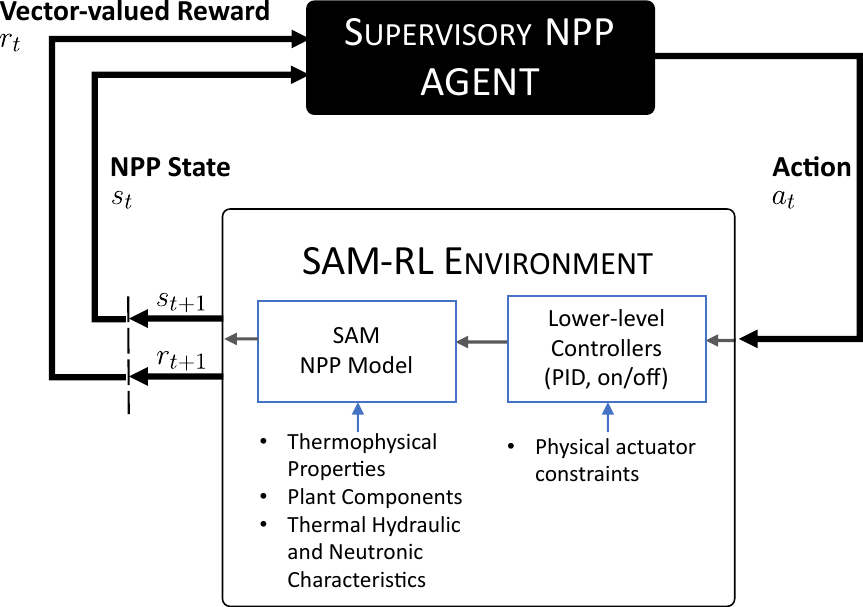}
\caption{Application of RL for supervisory NPP Control
using the proposed SAM-RL environment} 
\label{fig:SAM-RL}
\end{figure}

To design next-generation NPPs, significant efforts focus on developing advanced simulators.
The System Analysis Module (SAM) code is a state-of-the-art physics-based tool developed at Argonne National Laboratory to enable high-fidelity transient analyses of the whole plant for advanced NPP designs \cite{Hu2017}.
In this work, SAM is embedded into an RL environment as a digital twin to simulate the dynamics of the NPP.
The SAM-RL environment is presented in Figure \ref{fig:SAM-RL}.
The environment has two prominent features.
First, SAM will provide a physics-based simulation of NPPs by modeling heat transfer, fluid dynamics, and nuclear dynamics.
The SAM model is formed by the thermophysical properties of the system's fluid, the system's physical characteristics (e.g., coefficients defining dynamics of the nuclear reactor), and the specification and arrangement of physical components (e.g., pipes, heat exchangers, pumps).
Second, lower-level controllers will interface actions provided by the RL agent and the NPP's actuators.
This controller arrangement intentionally maintains overrideability and allows physical constraints to be applied to the actuators.

The RL agent interacts with the SAM-RL environment in discrete time steps $t=\{0,1,\cdots,T-1\}$ by exchanging system states $s_t$, actions $a_t$, and vector-valued rewards $r_t$.
Vector-valued rewards encode the primary~(\textit{e.g.}, load-following reward) and secondary objectives~(\textit{e.g.}, avoid constraint violation) of the controller. 
They can enforce physical constraints on system states or actuators as discussed in subsection \ref{subsec:prob-form}.

\begin{figure}[thbp]
    \centering
    \includegraphics[width=\columnwidth]{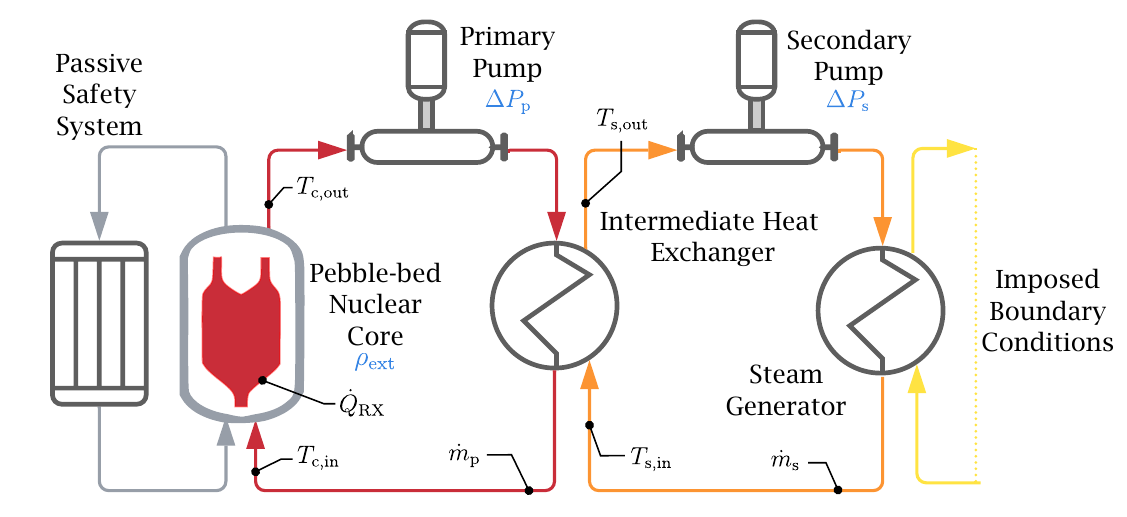}
    \caption{Layout of the Advanced NPP studied in this work.
        Measurement locations of various states are annotated.
        The major actuators are listed in \textcolor[HTML]{3584e4}{blue} text below the component labels.
    }
    \label{fig:npp-layout}
\end{figure}

In this work, we adopt a 320 MW pebble-bed fluoride-cooled high-temperature reactor design \cite{hu2020development}.
The layout of the power plant is presented in Figure \ref{fig:npp-layout}.
There are four circuits in total.
The primary side has a pump that circulates FLiBe molten salt through the intermediate heat exchanger and the nuclear core.
The secondary side has a pump that circulates FLiNaK molten salt through a steam generator and the intermediate heat exchanger.
Fixed heat transfer boundary conditions are imposed on the steam generator to mimic the presence of an energy conversion cycle.
Additionally, a passive safety system is present to aid in cooling the reactor vessel in the case of an accident.
This reactor design is modeled in SAM and then embedded in the SAM-RL environment.

\subsection{Control Strategy}\label{sec:controlstrat}

This work adopts a control strategy that reduces thermal cycling during load-follow operation \cite[\S 2.1]{dave2023design}.
The quantitative goal of this control strategy is to regulate the core inlet and outlet temperatures.
The reactor's power will be manipulated through external reactivity insertion to accommodate load increase or decrease requests.
Then, pumps on both the primary and secondary sides will govern the heat transfer rates across the primary heat exchanger.
The association between controlled and manipulated variables has been selected to optimize the impact of control actions and lessen cross-talk.
These associations are noted below:
\begin{itemize}
    \item The reactor power ($\qrx$) is controlled by external reactivity insertion ($\rhoext$).
    \item The core outlet temperature ($\tco$) is controlled by the primary pump head ($\dpp$).
    \item The core inlet temperature ($\tci$) is controlled by the solar salt pump head ($\dps$).
\end{itemize}
Three Proportional-Integral-Derivative (PID) controllers were designed, with two regulating the temperatures at the core outlet and inlet and the third tracking the reactor power.
The control strategy closely manages the dynamics of the primary loop (red circuit in \ref{fig:npp-layout}). 
Meanwhile, the secondary side of the plant is not directly controlled (orange circuit in \ref{fig:npp-layout}).
In \ref{fig:plant_performance}, the dashed lines show that the secondary-side inlet and outlet temperatures ($\tsi$ and $\tso$) achieve new equilibrium values once the load follows transient has completed.
To prevent temperature fluctuations and optimize the dynamics of the NPP, two constraints are set on the secondary side: a minimum temperature constraint on the heat exchanger inlet and a maximum temperature constraint on the heat exchanger outlet.

A common approach is to use a model that accurately represents the system dynamics to perform model predictive control. However, achieving the level of precision of the model for effective control is not trivial, and a different model must be established for each system. Model-free reinforcement learning algorithms relax this requirement by interacting with the environment and learning by trial and error.

\subsection{Reduced-order Environment}

Generating a transient of a power plant using SAM requires a computation time on the order of days. Training RL agents from scratch on SAM is, therefore, infeasible, as RL agents typically require hundreds of thousands of environment interactions. To avoid exhausting compute resources and speeding up the training of RL agents, we adopt a transfer learning approach whereby RL agents are first trained on a reduced-order model (ROM) of SAM. The ROM of SAM can be learned in a data-driven manner by generating $K$ transients in parallel using SAM and applying a sparse identification of the nonlinear dynamical system (SINDYc) \cite{brunton2016sparse} 
on the data. The learned ROM is used next to train the RL agents. \rv{To facilitate algorithm development and efficient training, we adopt a ROM sub-sampled every five steps from the original time scale (episode length $T=11250$), resulting in a shortened $T=2250$ for the experiments.  }

\section{Chance Constrained RL with Lagrangian Primal-Dual Optimization}\label{sec:method}

\subsection{Load-following Problem Formulation} \label{subsec:prob-form}

This work focuses on creating a Safe RL framework, which allows the agent to execute actions that optimize the cumulative rewards while adhering to state constraints. Load-following transient falls under this formulation, which 
is an important maneuver for advanced reactors to adopt due to the increasing proportion of intermittent energy sources. In the load-following transient, the task is to satisfy the requested changes in power generated by the operating unit while adhering to constraints placed on the system states. Prediction of violation is challenging, as the entire system has nonlinear behavior and lags associated with transport phenomena.

Mathematically, the problem can be formulated as a chance-constrained optimization problem
over the space of parameterized policies $\pi_{\vect{\theta}}(a|s)$ with parameters $\vect{\theta}$, which follows the safety level 1, constraint satisfaction encouraged, in the Safe RL context~\cite{brunkeSafeLearningRobotics2021}. 
The objective is to maximize the expected cumulative rewards subject to a chance constraint that ensures the system's state remains within a safe region over the operating horizon $T$ with a high probability,
\begin{equation} \label{eq:ccrl}
\begin{aligned}
&\underset{\pi}{\text{max}}~\mathbb{E}_{a_t \sim \pi, s_{t+1} \sim P, s_0 \sim \beta} \Big[ \sum_{t=0}^{T-1} \gamma^t \mathcal{R}(s_t,a_t) \Big] \\
& \text{s.t. } p_s(\pi) = \text{Pr}\Big[ \bigcap_{t=0}^{T-1} s_t \in \mathcal{S}_{\text{safe}} \Big] \geq 1-K\delta,
\end{aligned}
\end{equation}
\rv{where $\mathcal{S}_{\text{safe}}$ is the set of states that are within the safe operation range.}
In its current form, the optimization problem in \eqref{eq:ccrl} is hard to solve because the gradient of the chance constraint with respect to $\vect{\theta}$ is discontinuous, nondifferentiable, and crucially intractable. Notice that the Monte Carlo (MC) sample average to estimate $p_s(\pi)$ for a given $\pi$ is $\hat{p}_s(\pi) = \frac{1}{M} \sum_{m=0}^{M-1} \mathbbm{1}\big(\bigcap_{t=0}^{T-1} s_t^m \in \mathcal{S}_{\text{safe}}\big)$, where $s_t^m$ is the state of the system at time step $t$ in the $m$-th sampled trajectory, and $\mathbbm{1}\big(\bigcap_{t=0}^{T-1} s_t^m \in \mathcal{S}_{\text{safe}}\big)$ is an indicator function that is $1$ if the system remains within the safe region at all time steps $T$ and is $0$ otherwise. 

To solve equation \eqref{eq:ccrl} using model-free RL methods, we reformulate the chance constraint by applying Boolean algebra laws and probability bounds to arrive at a tractable form. First, notice that,
\begin{equation}
\resizebox{\columnwidth}{!}{$
\begin{aligned}
\text{Pr}\Bigg[ \overline{\bigcap_{t=0}^{T-1} s_t \in \mathcal{S}_{\text{safe}}} \Bigg] \leq K\delta
\Rightarrow  \text{Pr}\Big[ \bigcap_{t=0}^{T-1} s_t \in \mathcal{S}_{\text{safe}} \Big] \geq 1-K\delta,
\end{aligned}
$}
\end{equation}
that is, if we guarantee that the probability of the complementary event is at most $K\delta$, then the joint chance constraint will be satisfied with probability at least $1-K\delta$. By applying DeMorgan's law followed by a union bound,
\begin{equation} \label{eq:complement}
\begin{aligned}
\text{Pr}\Bigg[ \overline{\bigcap_{t=0}^{T-1} s_t \in \mathcal{S}_{\text{safe}}} \Bigg] &= \text{Pr}\Bigg[ \bigcup_{t=0}^{T-1} {s_t \not\in \mathcal{S}_{\text{safe}}} \Bigg]\\
& \leq \sum_{t=0}^{T-1} \text{Pr}\Big[{s_t \not\in \mathcal{S}_{\text{safe}}} \Big]
\end{aligned}
\end{equation}
In practice, the event ${s_t \in \mathcal{S}_{\text{safe}}}$ is usually expressed as $K$ (non)-linear constraint functions of the state, 
\begin{equation}
    {s_t \in \mathcal{S}_{\text{safe}}} \Leftrightarrow \mathcal{T}_k(s_t) \leq 0, \forall k \in \{1,\cdots, K\},
\end{equation}
where $\mathcal{T}_k$ is a (non)-linear operator on $s_t$. Define an indicator function, 
\begin{equation} \label{eq:indicator}
    \mathcal{C}_k(s_t) = 
    \begin{cases}
    0, \mathcal{T}_k(s_t) \leq 0\\
    1, \mathcal{T}_k(s_t) > 0,
    \end{cases}
\end{equation}
that is, the indicator function $\mathcal{C}_k(s_t)$ tracks whether or not the $k$-th safety constraint is violated at $s_t$. Notice that 
\begin{equation} \label{eq:not_safe}
\begin{aligned}
    \text{Pr}\Big[s_t \not \in \mathcal{S}_\text{safe}\Big] &\overset{(i)}{=} \text{Pr}\Big[ \bigcup_{k=1}^K \big(\mathcal{C}_k(s_t)=1\big) \Big] \\&\overset{(ii)}{\leq} \sum_{k=1}^K\text{Pr}\Big[\mathcal{C}_k(s_t)=1 \Big],
\end{aligned}
\end{equation}
where $(i)$ again follows by applying DeMorgan's law, and $(ii)$ follows by applying a union bound. Combining the results of \eqref{eq:complement} with that of \eqref{eq:not_safe}, we obtain the following implications, 
\begin{equation}
    \begin{aligned}
    \sum_{t=0}^{T-1} \text{Pr}\Big[{s_t \not\in \mathcal{S}_{\text{safe}}} \Big] &\leq \sum_{t=0}^{T-1}\sum_{k=1}^K\text{Pr}\Big[\mathcal{C}_k(s_t)=1 \Big] \leq K\delta \\
    &\Rightarrow p_s(\pi) \geq 1-K\delta.
    \end{aligned}
\end{equation}
It is also true that if $\sum_{t=0}^{T-1}\text{Pr}\Big[\mathcal{C}_k(s_t)=1 \Big] \leq \delta, \forall k \in \{1, \cdots, K\}$, the original chance constraint will be satisfied with probability at least $1-K\delta$. Notice that, 
\begin{equation}
\begin{aligned}
    \mathbb{E}\Big[\sum_{t=0}^{T-1} &\mathcal{C}_k(s_t) \Big] = \sum_{t=0}^{T-1} \mathbb{E}\Big[\mathcal{C}_k(s_t) \Big] \\
    &= \sum_{t=0}^{T-1} \text{Pr}\Big[\mathcal{C}_k(s_t)=1 \Big]*1 + \text{Pr}\Big[\mathcal{C}_k(s_t)=0 \Big]*0 \\
    &=\sum_{t=0}^{T-1} \text{Pr}\Big[\mathcal{C}_k(s_t)=1 \Big]\\
\end{aligned}
\end{equation}
Furthermore, $\mathbb{E}\Big[\sum_{t=0}^{T-1} \mathcal{C}_k(s_t) \Big] \leq \frac{1}{1-\gamma} \mathbb{E}\Big[\sum_{t=0}^{T-1} \gamma^t \mathcal{C}_k(s_t) \Big]$ as long as $\frac{\gamma^t}{1-\gamma} \geq 1, \forall t \in \{0,\cdots,T-1\}$ (e.g. for $T=2250, \gamma \gtrapprox 0.9974$). Hence, a more conservative yet tractable reformulation of \eqref{eq:ccrl} is, 

\begin{equation} \label{eq:ccrl2}
\resizebox{\columnwidth}{!}{$
\begin{aligned}
&\underset{\pi}{\text{max}}~\mathbb{E}\Big[ \sum_{t=0}^{T-1} \gamma^t \mathcal{R}(s_t,a_t) \Big] \\
& \text{s.t.}~\mathbb{E}\Big[\sum_{t=0}^{T-1} \gamma^t \mathcal{C}_k(s_t) \Big] \leq \delta(1-\gamma), \forall k \in \{1,\cdots, K\}.
\end{aligned}
$}
\end{equation}

By learning the optimal policy for \eqref{eq:ccrl2}, we ensure that this policy is feasible, albeit being over-conservative, with respect to \eqref{eq:ccrl}, i.e. $p_s(\pi)\geq 1-K\delta$. 
The optimization problem of \eqref{eq:ccrl2} can be readily formulated as a discrete-time cMDP with continuous state-action spaces as follows, 
\begin{enumerate}
    \item $\forall s_t \in \mathcal{S}, s_t = [\vect{x}_t,\vect{c}_t,p_t] \in \mathbb{R}^d$, where $\vect{x}_t$ is a vector denoting the state of the power plant at time step $t$, $\vect{c}_t$ is a vector denoting NPP state constraint bounds, and $p_t$ is a  scalar that denotes the power demand,
    \item $\forall a_t \in \mathcal{A}, a_t \in \mathbb{R}^+$ is the action that denotes the power output from the plant,
    \item $r_t = [r_t^0,r_t^1,\cdots,r_t^K] \in \mathbb{R}^{K+1}$ is a vector valued reward function which encodes the primary $(r_t^0)$ and $K$-secondary objectives $(r_t^1,\cdots,r_t^K)$ of the RL-based controller. $r_t^0$ is the negative of the squared $\text{L}2$ norm between the power demand $p_t$ and power plant supply $a_t$, i.e. $r_t^0 = \mathcal{R}(s_t,a_t)= -||p_t - a_t||_2^2$. By maximizing a cumulative sum of the primary reward signals, the agent is trained to learn a policy that satisfies the power demand as much as possible. On the other hand, the $k$-th secondary objectives $r_t^k=\mathcal{C}_k(s_t)$ is the indicator function that is $1$ if $s_t$ violates the $k$-th constraint on the state of the system and is $0$ otherwise, as given by Equation \eqref{eq:indicator}.
\end{enumerate}

\subsection{Lagrangian Primal-Dual Optimization}
In this subsection, we design an on-policy RL algorithm to solve \eqref{eq:ccrl2}. The on-policy algorithm is based on proximal policy optimization (PPO)~\cite{schulman2017proximal}. PPO has shown the advantages of its scalability, data efficiency, and robustness in various applications~\cite{ouyang2022training, kiran2021deep}. To incorporate chance-constrained state optimization under the framework of PPO, we leverage the idea of Lagrangian relaxation in which \eqref{eq:ccrl2} is converted to an equivalent unconstrained problem, 

\begin{equation} \label{eq:ccrl3}
\resizebox{1\columnwidth}{!}{$
\begin{aligned}
&\underset{\pi}{\text{max}}~\underset{\{\lambda_k\geq0\}}{\text{min}}~\mathbb{E}\Big[ \sum_{t=0}^{T-1} \gamma^t \Big( \mathcal{R}(s_t,a_t) -\sum_{k=1}^{K}\lambda_k \mathcal{C}_k(s_t)\Big)\Big] + \sum_{k=1}^{K}\lambda_k\delta(1-\gamma) ,
\end{aligned}$}
\end{equation}

and invoking the min-max theorem to interchange the max and min terms, 

\begin{equation} \label{eq:ccrl4}
\resizebox{1\columnwidth}{!}{$
\begin{aligned}
&\underset{\{\lambda_k\geq0\}}{\text{min}}~\underset{\pi}{\text{max}}~\mathbb{E}\Big[ \sum_{t=0}^{T-1} \gamma^t \Big( \mathcal{R}(s_t,a_t) -\sum_{k=1}^{K}\lambda_k \mathcal{C}_k(s_t)\Big)\Big] + \sum_{k=1}^{K}\lambda_k\delta(1-\gamma) ,
\end{aligned}$}
\end{equation}

which can be solved on two-time scales: on a faster time scale, gradient ascent is performed on state values to find the optimal policy for a given set of Lagrangian variables, and on a slower time scale, gradient descent is performed on dual variables \cite{liang2018accelerated}. Notice that for a \textit{fixed} set of Lagrange variables, the problem reduces to a standard MDP with a penalized reward function \cite{tessler2018reward},

\begin{equation} \label{eq:penalized_reward}
    \hat{\mathcal{R}}(s_t,a_t) =\mathcal{R}(s_t,a_t) -\sum_{k=1}^{K}\lambda_k\mathcal{C}_k(s_t) +\text{constant}.
\end{equation}
where $\text{constant}=\sum_{k=1}^{K}\lambda_k{\delta(1-\gamma)}$ is a constant \rv{for policy and value network updates at the faster time scale},  thus can be dropped from the optimization objective.
Optimization time scales are controlled by choosing the maximum learning rate of the stochastic gradient optimizer used, e.g., adaptive moment estimation \cite{kingma2017adam}, and the number of gradient steps performed at the end of each training epoch.

\subsection{Constrained PPO}

We use PPO to train an RL agent to maximize the proposed objective~(\ref{eq:penalized_reward}). PPO is a state-of-the-art on-policy RL algorithm in which the parameterized policy $\pi_{\vect{\theta}}(a|s)$ is directly learned by maximizing the PPO-clip objective function,
\begin{equation} \label{ppo}
\resizebox{1\hsize}{!}{$
\mathcal{O}^\text{clip}(\vect{\theta}) = {\mathbb{E}}_t\Big[\text{min}( \frac{\pi_{\vect{\theta}} (a_t|s_t)}{\pi_{\vect{\theta}_\text{old}}(a_t|s_t)} \hat{A}_t, \text{clip}(\frac{\pi_{\vect{\theta}} (a_t|s_t)}{\pi_{\vect{\theta}_\text{old}}(a_t|s_t)}, 1+\epsilon, 1-\epsilon)\hat{A}_t) \Big]$},
\end{equation}
where $\vect{\theta}$ are the policy parameters, $\epsilon$ is a clip fraction, and $\hat{A}_t$ is the generalized advantage estimator (GAE) \cite{schulman2015high}
, 
\begin{equation} \label{gae}
    \hat{A}_t = \sum_{l=0}^\infty (\gamma \xi)^l \big(\hat{\mathcal{R}}_{t+l} + \gamma V_{\vect{\phi}}(s_{t+l+1}) - V_{\vect{\phi}}(s_{t+l}) \big),
\end{equation}
where $\xi$ is a hyper-parameter to strike a balance between bias and variance in estimating the advantage, and $V_{\vect{\phi}}(s_t)$ is the estimated parameterized value function with parameters $\vect{\phi}$. The clipped surrogate advantage objective \eqref{ppo} ensures that gradient-based policy updates yield improved policies by limiting (by clipping) how much the new policy can move away from the old policy while improving the training objective. To further constrain policy updates, the gradient optimization subroutine running on a batch of trajectories is typically terminated when the expected KL-divergence between the new policy and the old policy reaches a predefined threshold $\text{KL}_{\text{Th}}$.

In \eqref{gae}, $\hat{\mathcal{R}}_{n}$ is the Lagrangian penalized reward function given by \eqref{eq:penalized_reward}. The Lagrangian penalty multipliers are updated according to policy feasibility by gradient descent on the original constraints. Given that $\lambda_k,\forall k$ are initially set to $0$, i.e., the agent is initially indifferent to the cost constraints, the Lagrangian penalty multipliers are updated at a slower time scale than that of updating the policy by minimizing the following loss function with respect to $\lambda_k,\forall k$,
\begin{equation} \label{lag:loss}
\resizebox{1\hsize}{!}{$
    \mathcal{O}^{L}(\lambda_1,\cdots,\lambda_K)=\sum_{k=1}^K \lambda_k \text{clip}\Big({\delta(1-\gamma)} - \mathbb{E} \Big[\sum_{t=0}^{T-1} \gamma^t \mathcal{C}_k(s_t) \Big], -\infty, 0 \Big)$}
\end{equation}
As long as a constraint $k$ is violated, then $\frac{\partial \mathcal{O}^{L}}{\partial \lambda_k} < 0$, and so $\lambda_k$ will be increased to enforce the constraint. Due to clipping, $\lambda_k$ will not be updated if the constraint $k$ is satisfied. By restricting $\lambda_k,\forall k$ to increase monotonously during training, the agent avoids oscillations between feasible and infeasible policy spaces when optimizing policy parameters \eqref{ppo}, which improves learning stability and promotes convergence to a local saddle point.

Finally, the state-value function is learned by minimizing the mean squared error loss against the policy's discounted penalized rewards-to-go.
\begin{equation} \label{mse:V}
    \mathcal{O}^\text{V}(\vect{\phi}) = \mathbb{E}\Big[ \Big(V_{\vect{\phi}}(s_t) - \sum_{l=0}^\infty \gamma^l \hat{\mathcal{R}}_{t+l}(s_{t+l},a_{t+l}) \Big)^2 \Big].
\end{equation}

\subsection{RL Agent Training}
\subsubsection{Policy/Value Network Architecture}

The policy is a parameterized Gaussian policy, shown in Equation \ref{eqn:policy}. During training, we sampled the action from the Gaussian policy and completed episodes. An invertible squashing function $tanh$ was applied to the policy to enforce action bounds. The distribution of the squashed policy and its log-likelihood can be easily derived as in \cite{haarnoja2018soft}. 

\begin{equation}
    \pi(a_t|s_t, \vect{\theta_\pi}) = \frac{1}{\sigma(s_t,\vect{\theta_\pi})\sqrt{2\pi}} \text{exp} \Big(- \frac{(a_t - \mu(s_t,\vect{\theta_\pi}))^2}{2\sigma(s_t,\vect{\theta_\pi})^2} \Big).
    \label{eqn:policy}
\end{equation}

We parallelize the training process with multiple workers through MPI and randomly sample a demand curve for each worker per epoch. During training, for each episode, the policy network inputs the state variables (primary-side mass flow rate, secondary-side mass flow rate, core inlet temperature, core outlet temperature, core outlet pressure, secondary-side heat-exchanger inlet temperature, secondary-side heat-exchanger outlet temperature, delayed neutron precursor concentrations, total heat-exchanger energy transfer, and energy generated by the steam generator), demand, and inlet/outlet temperature constraints, and outputs the Gaussian policy's mean and standard deviation functions. The SINDYc environment then takes a sampled action from the policy and returns the state variables and demand for the next time step until it reaches the end of the episode. The value networks approximate the value of current and future states. Finally, the model calculates the reward values and updates the model parameter via gradient ascent over the reward function on the faster time scale and gradient descent to update the Lagrangians on a slower time scale. We control the randomization of the generated power demand curves and designate 36$\times$600  unique episodes for training, 10 for validation and 30 for testing.

\section{Results}
\begin{figure*}[thbp]
    \centering
    \includegraphics[width=\linewidth]{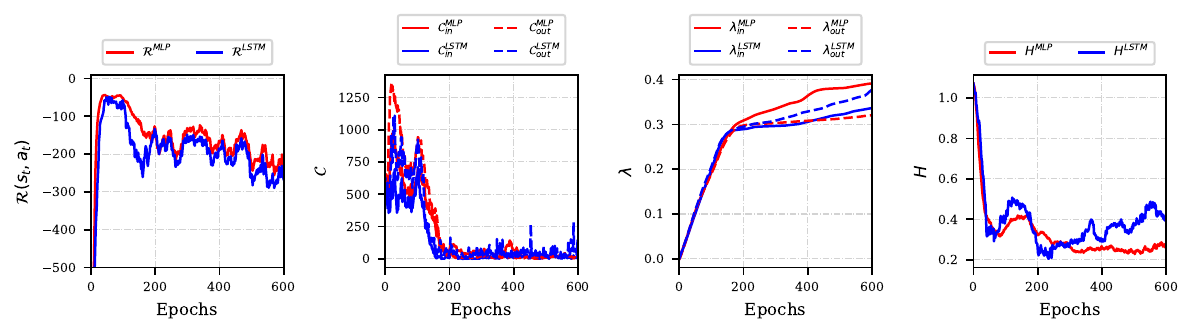}
    \caption{The training curves of the proposed models with MLP- and LSTM-actors. From left to right, it shows the changes in rewards, costs associated with two safety constraints, the magnitude of learned Lagrangian multipliers, and the policy entropy.}
    \label{fig:modelTraining}
\end{figure*}
This section shows the experimental results using the proposed model for performing load-following in generic fluoride high-temperature reactors. The model evaluation is based on reporting three metrics scores. We first show the training and testing results of the proposed models using multi-layer perceptron~(MLP) and Long-short term memory~(LSTM) actors, respectively. 
Furthermore, we compare the proposed model with other baselines where we fix the Lagrangian multipliers in (\ref{eq:penalized_reward}) as a regularization term to the load-following reward function, which can be regarded as a means of reward shaping~\cite{laud2004theory}.   Finally, we deploy the trained $\lambda$-PPO agent to the full-fidelity SAM model with action clipping for stability and show control results of the direct application of the trained agent.

The implementation and models were in TensorFlow with an MPI setup, and the training and evaluation were on the Bebop cluster with 36 workers from Argonne's Laboratory Resource and Computing Center~(LCRC).

\subsection{Evaluation Metrics}
We use three metrics to evaluate model performance, the reward-cost score, $\mathcal{\bar{R}}(s_t,a_t)$, the mean distance of violation, $D$, and the mean proportion of violation or violation rate, $\omega$. 

$\mathcal{\bar{R}}(s_t,a_t)$, defined in (\ref{eqn:reward-cost}), is simply the load-following reward minus the cost values for violating the constraints. A higher value indicates better performance. 

\begin{equation}\label{eqn:reward-cost}
	\mathcal{\bar{R}}(s_t, a_t) = \mathcal{R}(s_t, a_t) - \sum_{k=1}^2\mathcal{C}_k(s_t, a_t).
\end{equation}
Since the violation costs are binary values, that is, violations of constraints contribute a value of 1 and 0; otherwise, costs do not directly capture the magnitude of the violation a model causes. Therefore, to account for the magnitude of violation, we also introduce the mean distance of violation, $D$, defined in (\ref{eq:distanceViolation}).
\begin{equation}\label{eq:distanceViolation}
\begin{aligned}
	D_{in} &= \sum_{i=1}^N \mathbb{I}(\mathcal{C}_{in} - T_{in})\mid T_{in} - \mathcal{C}_{in} \mid, \\
	D_{out} &= \sum_{i=1}^N \mathbb{I}(T_{out} - \mathcal{C}_{in})\mid T_{out} - \mathcal{C}_{out} \mid, \\
	D &= \frac{D_{in} + D_{out}}{2N},
\end{aligned}
\end{equation}
where $N$ is the number of trajectories in the testing set, and $\mathbb{I}$ is an indicator function~(\ref{eq:indicator}), which takes value 1 when there is a violation of constraints.
\begin{equation}\label{eq:indicator}
	\mathbb{I}(x) = \begin{cases}
		0 & \text{if}~x \leqslant 0\\
		1 & \text{if}~x > 0
	\end{cases}.
\end{equation}

The third metric is the mean violation rate defined as (\ref{eq:VR}),

\begin{equation}\label{eq:VR}
	\omega = \frac{1}{N} \sum_{i=1}^N \frac{t^{(i)}_{v}}{
 \tau^{(i)}},
\end{equation}

where $t^{(i)}_{v}$ is the time duration during which a constraint is violated, and $\tau^{(i)}$ is the total duration of the $i$th trajectory.

We use all three metrics to evaluate the models in the following sections. A lower number indicates better model performance for all these metrics.


\subsection{Load-following in generic fluoride high-temperature reactors (gFHR) }

 We trained the proposed chance-constrained PPO agent on $36\times 600$~ (600 epochs and 36 workers per epoch) \textit{unique} episodes and validated it using 10 other unique episodes. The objective of the model was to learn a policy that maximizes the cumulative reward over each episode, ensuring that the inlet/outlet temperature does not breach the time-dependent constraints. 
 We used an MLP-based and LSTM-based actor to further analyze the impact of the policy network structure on the agent's performance.

\begin{figure*}[t]
    \centering
    \includegraphics[width=\linewidth]{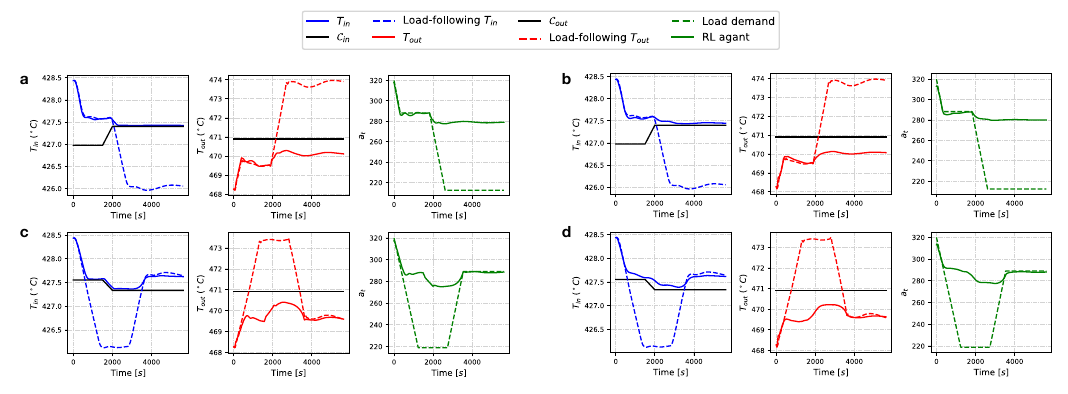}
    \caption{Visualization of agents' performance on a testing trajectory. \textbf{a.} \& \textbf{c.} show the model performance of the MLP actor; \textbf{b.} \& \textbf{d.} show the model performance of the LSTM actor. Both models resulted in compliance with safety constraints while closely following demand when possible. The MLP actor, in comparison, adhered more closely to the demand and safety constraints, whereas the LSTM actor led to fewer oscillations in actions. In \textbf{c.} \& \textbf{d.}, both trained RL agents adapt successfully to time-varying constraints. The resulting states adhere closely to the changing constraints and return to following the demand-induced trajectory when potential violations no longer present.}
    \label{fig:agentPerformance}
\end{figure*}

\begin{figure*}[thbp]
	\centering
	\includegraphics[width=0.9\linewidth]{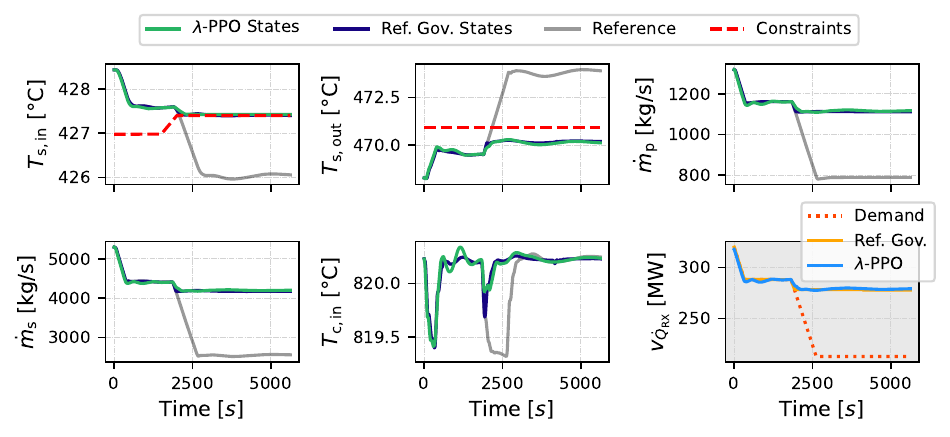}
	\caption{The visualization of the response of other state variables to load demand and $\lambda$-PPO agent. Compared to the reference governor, $\lambda$-PPO agent adheres to the constraints more closely without violation.} 
    \label{fig:plant_performance}
\end{figure*}

We trained MLP- and LSTM-actor PPO agents on the same training trajectories. Figure \ref{fig:modelTraining} shows the training dynamics of the proposed models. Both models present a drastic jump in reward at the beginning of training and a gradual decrease due to the enforcement of safety constraints as training progresses. As a result, the corresponding costs $\mathcal{C}_{in}$ and $\mathcal{C}_{out}$, decrease, suggesting that agents are learning policies that follow demand while staying in safe regions. The trainable Lagrangian multipliers, $\lambda_{in}$ and $\lambda_{out}$, in both MLP-actor and LSTM-actor models, increase sharply at the early stage of training when the agents simply try to follow the demand only. At a later time, the rate of increase of $\lambda_{in}$ and $\lambda_{out}$ slows down as the agents learn to satisfy the constraints simultaneously. The entropy of the Gaussian policy implies the agent's exploration rate. Sufficient exploration is necessary for the agent to obtain different experiences to find the optimal policy. However, excessive exploration can diverge from optimal policy and make recovery difficult. In training the proposed models, the policy entropy started with high values, ensuring abundant exploration, and gradually decreased to exploit the best possible policy.



In comparison, the MLP actor model showed a higher load-following reward, lower constraint costs, and lower entropy at the end of training. The learned $\lambda_{in}$ and $\lambda_{out}$ of the MLP-actor lie between them of the LSTM-actor, which suggests a harsher regularization for $T_{in}$ and less for $T_{out}$ in the MLP-actor. Regarding the policy entropy, the MLP-actor ended up with a stabilized value of around 0.2 while the LSTM-actor presents a more oscillatory and higher policy entropy with a value of around 0.4. To evaluate the performance of the RL model, we calculated $\mathcal{\bar{R}}$, $D$, and $VR$~(Table \ref{tab:MLP-LSTM}) over 30 testing episodes and visualized several trajectories. We compared the state variable trajectories controlled by the RL agent with the trajectories obtained by following the demands. Figure \ref{fig:agentPerformance} shows that the actions controlled by the RL agent lead to non-violation of the constraints while closely following the demand where possible. \rv{In comparison, the inlet temperature controlled by the reference governor is slightly lower than the constraint lower bound. Meanwhile, the expected states from simply following the demand evidently violate both upper and lower bounds of the constraints.} 

\begin{table}
	\centering
	\caption{Model performance of the proposed models}
	\begin{tabular}{lcccc}
	\toprule
	Models   & $\bar{\mathcal{R}}$ & $D$ & $\omega$ \\
	\midrule
	MLP-actor & $-251.84$ &$3.2657 \times 10^{-5}$ & 0.0015\\
	LSTM-actor & $-450.27$ & 0.0016 & 0.0796\\
	\bottomrule	 
	\end{tabular}\label{tab:MLP-LSTM}
\end{table}


The performance difference between the MLP and LSTM actors was because, on the one hand, the system was Markovian, where the states at the next step only depend on the current states. The MLP actor depended only on the current-state input and the output of the corresponding action. This led to a faster response and closer adherence to the constraints of the MLP actor.  On the other hand, the LSTM actor considered the hidden states in the recurrent layer, of which the weights were trained on a certain length of experience history~(trace length). Therefore, the LSTM actor resulted in a slower response to constraint changes and a smoother action output.

Given the superior performance of the MLP-actor PPO agent, we call it $\lambda$-PPO model
and use it for the subsequent results. 
Fig.~\ref{fig:plant_performance} shows the performance of the trained PPO agent.
The agent is given the task of reducing the power of the plant through several ramps, shown in the $v_{\dot{Q}_{RX}}$ subplot.
When reducing power, the agent is required to adhere to constraints on the inlet and outlet of the secondary heat exchanger ($T_\mathrm{s,in}$ and $T_\mathrm{s,out}$).
Due to the control strategy, Section \ref{sec:controlstrat}, the temperatures at the primary circuit are regulated (\textit{i.e.}, to be kept constant).
The secondary side settles to a new value, while the regulated temperatures revert to the initial values once the power change is completed.
The agent attempts to meet the requested load change while satisfying the constraints on the changes in secondary-side temperatures.
At the beginning of the transient, the agent correctly follows the requested load because no constraints are anticipated to be violated.
However, at around 2100 s, a constraint violation would occur if the agent continues to follow the reduction in load.
The agent correctly intervenes to avoid a constraint violation while attempting to meet the reduction in load.

\begin{figure}[thbp]
    \centering
    \includegraphics[width=.9\linewidth]{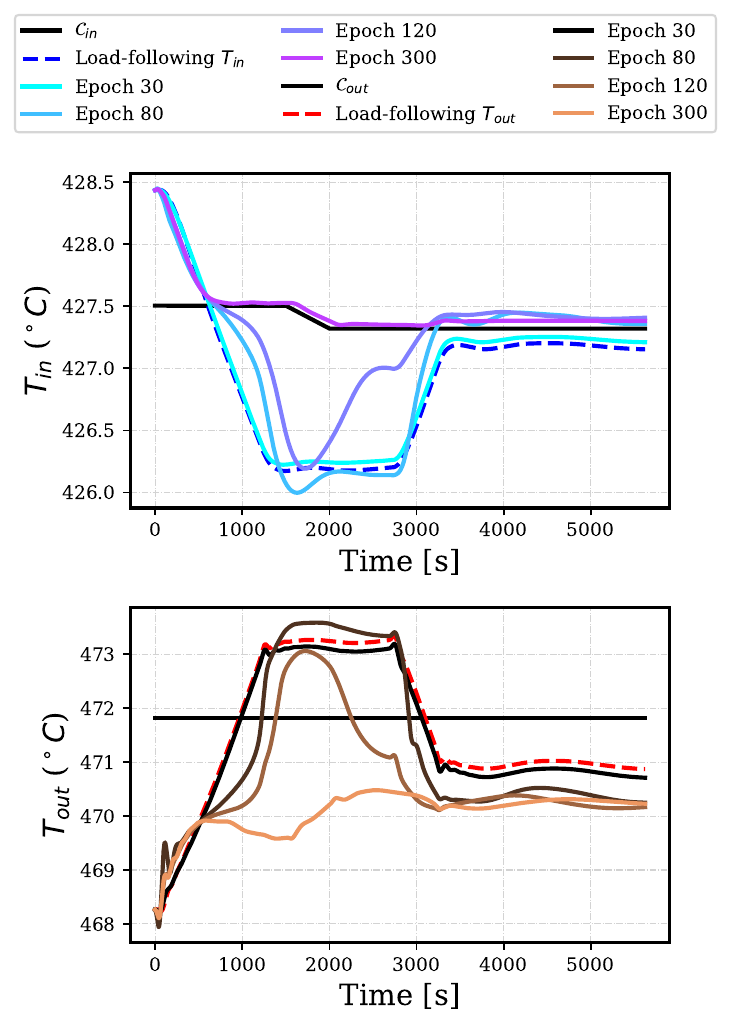}
    \caption{Agent performance controlling the inlet and outlet temperature during training. The magnitude of the Lagrangian multipliers increased as the training went on. At the initial stage of training, the model closely follows the demand, violating the constraints. However, as the number of epochs increased, the model acted within safety constraints while following the demand when it was feasible.}
    \label{fig:varyLags}
\end{figure}

Instead of fixing the Lagrangian multipliers in (\ref{eq:penalized_reward}), learning is essential to maximize the possible cumulative reward while satisfying constraints. The learned Lagrangian multipliers effectively regulate the action of the agent to satisfy the safety constraints. Figure~\ref{fig:varyLags} shows the progression of the Lagrangian effect during training as they gradually increase. At the start of training, as updating the Lagrangians was on a slower time scale, the policy was merely to follow the demand without safety concerns. After a few epochs, the learned Lagrangians increased, resulting in better compliance with the constraints.

\begin{figure*}[thbp]
	\centering
	\includegraphics[width=.9\linewidth]{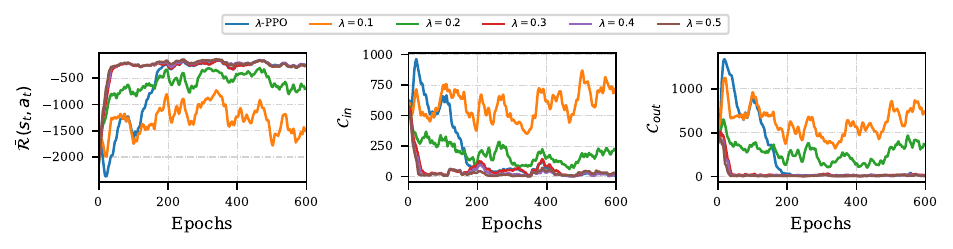}
	\caption{Learning dynamics of the proposed $\lambda$-PPO and other baseline models.}\label{fig:baselines}
\end{figure*}

\begin{table}[h]
	\centering
	\caption{Model performance comparison}
		\label{tab:baselines}
	\begin{tabular}{lcccc}
	\toprule
	Models & $\bar{\mathcal{R}}$ &  $D$ & $\omega$\\
	\midrule
	$\lambda$-PPO &$\mathbf{-251.84}$& $\mathbf{3.2658 \times 10^{-5}}$ &\textbf{ 0.0015}\\
	$\lambda = 0.1$ &$-1600.28$ & 0.6652 & 0.3698\\	
	$\lambda = 0.2$ & $-746.00$& 0.2611&0.0923 \\	
	$\lambda = 0.3$ &$-261.85$ &$5.1481 \times 10^{-5}$&0.0039 \\	
	$\lambda = 0.4$ &$-258.42$ &$3.5640 \times 10^{-5}$ & 0.0033 \\	
	$\lambda = 0.5$ & $-273.03$ &$9.5473 \times 10^{-5}$& 0.0049\\	
	\bottomrule
	\end{tabular}

\end{table}

\subsection{Learned Lagrangians}
We compared the performance of our proposed model, which we call $\lambda$-PPO, with baseline models with fixed Lagrangian multipliers. Baseline models can be viewed following a reward-shaping scheme~\cite{laud2004theory}, where we explicitly impose constraints on the objective function. Figure~\ref{fig:baselines} and Table~\ref{tab:baselines} show the comparison of these models in terms of $\mathcal{\bar{R}}$, $D$ and $\omega$. The results suggest a distinctly superior performance of $\lambda$-PPO with the lowest violation distance $D$ and violation rate $VR$. Meanwhile, with a small fixed $\lambda$~(\textit{e.g.} 0.1 and 0.2),  the agent resulted in a low reward-cost score, $\mathcal{\bar{R}}$, and high constraints violation costs, $\mathcal{C}_{in}$ and $\mathcal{C}_{out}$. With a larger fixed $\lambda$ (\textit{e.g.} 0.4 and 0.5), which happened to be close to the learned values,  the agent gained an evident improvement in satisfying the constraints. However, the degree of violation of the constraint and the violation rate are higher than the $\lambda$-PPO model. The proposed $\lambda$-PPO model was shown to have the ability to learn the most appropriate Lagrangian multipliers in (\ref{eq:penalized_reward}) from the given environment, eliminating the extra effort of manual fine-tuning.

\subsection{Transfer Learning on the full SAM model}



\begin{figure*}[thbp]
    \centering
    \includegraphics[width=0.8\linewidth]{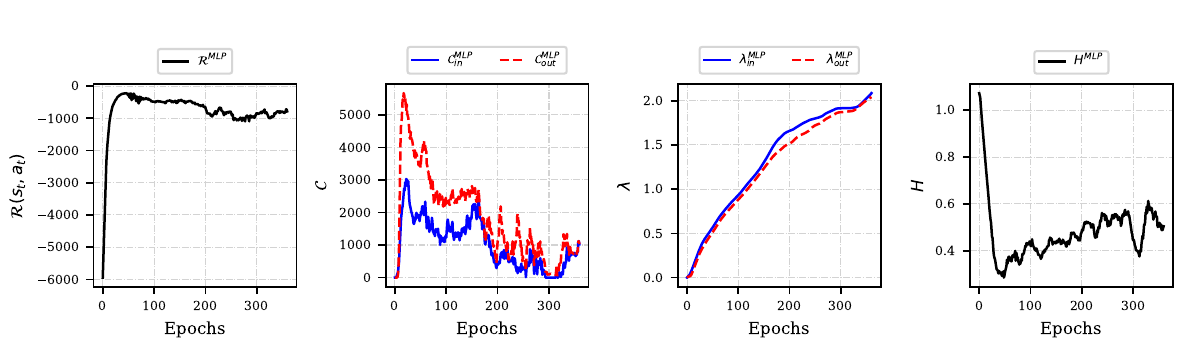}
    \caption{Learning curve for MLP $\lambda$-PPO agent on fine resolution reduced-order SINDYc environment.}
    \label{fig:fine_train_curve}
\end{figure*}

Given the fact that the SAM model needs to run on the original time resolution,
we trained a new $\lambda$-PPO agent in the reduced-order SINDYc environment with fine time resolution. Due to the much longer episode length compared to the previous environment (11250 vs. 2250), to reduce the variance in value and advantage function calculation, the model treated a long episode as 5 shorter sub-episodes~(same length as the previous episode). The bootstrapped values of the value network output replaced the values of the last state in the sub-episodes. Figure~\ref{fig:fine_train_curve} shows the training dynamics of such $\lambda$-PPO agent.

\begin{figure}
    \centering
    \includegraphics[width=\linewidth]{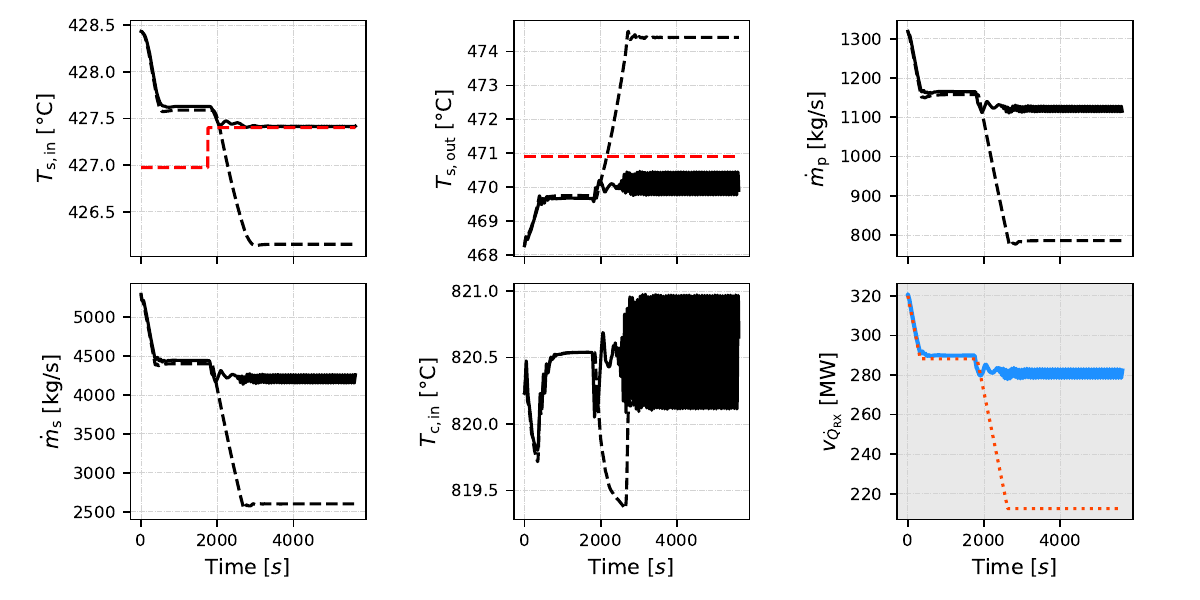}
    \caption{$\lambda$-PPO agent's performance as a result of direct application to the full-fidelity SAM environment.}
    \label{fig:sam_rollout}
\end{figure}

\textit{Direct application of the agent.} We investigated the agent's performance on controlling the high-fidelity SAM model trained using the reduced-order SINDYc environment. The SINDYc environment is a linear approximation of the full-fidelity SAM environment, hence the changes in actions need to be further constrained to prevent instability in the system. As a result, at rollout time, the RL agent followed a designed action clipping scheme. The action clipping is as follows.

\begin{equation}
    \begin{aligned}
        a_t &\leftarrow a_t + \Delta a_{t, t-1}\\
        \Delta a_{t, t-1} &= \text{clip}(-\eta, \eta, a_t - a_{t-1})
    \end{aligned}
\end{equation}
In this particular experiment $\eta = 5 \times 10^{-4}$, determined by the model performance on the validation trajectories from SINDYc environment. Clipped actions mitigate oscillations that could cause instability while allowing the agent to follow the demand closely. Figure \ref{fig:sam_rollout} shows the $\lambda$-PPO agent's performance in full-fidelity SAM environment 

In the initial phase of the trajectory, the $\lambda$-PPO agent adheres closely to the demand. However, when the state $T_\mathrm{s,in}$ approaches violation, the agent ceases to follow the demand, leading to a transient phase characterized by oscillations in the agent's actions. These oscillations subsequently induce instability in other states.
This unstable behavior can be traced back to the contrasting dynamics between two environments: the SINDYc and SAM. In the SINDYc environment, a surrogate model is trained on data that includes load demand curves but excludes any direct interventions or abrupt changes from an agent. Conversely, the SAM environment is a high-fidelity, physics-based model where multiple coupled Partial Differential Equations (PDEs) are solved to simulate the system's behavior.
A key equation in SAM is the point-kinetics equation (PKE), which governs the propagation of power within a Nuclear Power Plant. In this context, power increases typically follow an exponential pattern. Consequently, any sharp alterations in reactivity insertion (the variable controlling reactor power) can lead to exponential changes in the generated power.
We hypothesize that the SINDYc model does not capture this nuanced relationship between input and reactor power, as it is trained only on data with ``smooth'' changes. This limitation manifests in the agent's behavior when trained on the SINDYc environment, resulting in small oscillations in the power setpoint as it attempts to reconcile the conflicting demands of load and temperature constraints.
To rectify this behavior, future research will explore transfer learning techniques, allowing the SINDYc-trained agent to adapt and learn within the more complex SAM environment. This approach aims to bridge the gap between the simplified dynamics of the SINDYc model and the intricate, real-world physics represented in the SAM environment.

\section{Conclusion}
We proposed $\lambda$-PPO model for applying principles of Safe RL to the control of a complex dynamical system. We formulated Safe RL problem as a chance-constrained Markov Decision Process and transformed the intractable original form into an unconstrained standard MDP with the primal-dual Lagrangian method. As a result, we proposed $\lambda$-PPO model that adopts Proximal Policy Optimization to train agents in an on-policy manner. The model training followed two timescales - the policy and value networks were updated on a faster scale with a fixed set of Lagrangian multipliers, while the Lagrangians were learned on a slower timescale given the current policy. We applied the proposed model to a load-following transient problem where the agent was trained and evaluated using reduced-order SINDYc models. The results show that the $\lambda$-PPO agent followed the load demand and adhered to the temperature restrictions when following the demand could result in safety violations. Furthermore, we demonstrated the superiority of learning Lagrangians over fixing them to constant values, where the agent with the learned Lagrangians had the best performance regarding the distance from the violation, $D$, and the violation rate $\omega$. 

The application of the trained model to full-fidelity SAM suggests its efficacy in real-world scenarios. However, due to the lost nonlinearity in the reduced-order SINDYc environments, the agent's actions resulted in oscillations with different degrees in some states. Therefore, in future work, we propose to address this issue by 1) building a nonlinear ROM for SAM using physics-informed neural networks and 2) offline training of RL models that can benefit from existing data collected from the SAM model that contains the nonlinear information.

\section*{Acknowledgment}
This work is supported by Argonne National Laboratory under the Laboratory Directed Research \& Development project 2022-0077.



\bibliographystyle{elsarticle-num}
\bibliography{references}

\end{document}